\documentstyle[preprint]{ptptex}

\title{
Note on the Deformed Boson Scheme in Four Kinds of Boson Operators
}

\author{
Atsushi {\sc Kuriyama},$^{1}$ 
Constan\c{c}a {\sc Provid\^encia}$^{2}$, \\
Jo\~ao da {\sc Provid\^encia}$^{2}$, Yasuhiko {\sc Tsue}$^{3}$ 
and Masatoshi {\sc Yamamura}$^{1}$
}

\inst{
$^1$Faculty of Engineering, Kansai University, Suita 564-8680, Japan\\
$^{2}$Departamento de Fisica, Universidade de Coimbra, P-3000 Coimbra, 
Portugal\\
$^{3}$Physics Division, Faculty of Science, Kochi University, Kochi 780-8520, 
Japan
}

\recdate{
}

\abst{
The deformed boson scheme in four kinds of boson operators, 
which was recently proposed by the present authors, is supplemented 
by the $T$-type deformation closely related with the $su(1,1)$-algebra. 
Two subjects are discussed in relation to the $S$-type 
deformation closely related with the $su(2)$-algebra. 
}

\begin{document}

\maketitle


In the previous paper,\cite{KPPTYsub}
which is hereafter referred to as (I), the present authors proposed 
the deformed boson scheme for many-body systems in four kinds of 
boson operators. The basic idea in (I) comes from a series of 
three papers by the present authors.\cite{KPPTY01-1,KPPTY01-2,KPPTY02}
A coherent state, which belongs to the orthogonal set for boson system 
under investigation, is the most fundamental element for our deformed boson 
scheme. This scheme is obtained by deforming the coherent state. 
As was shown in Ref.\citen{KPY00}, the orthogonal set for four kinds of 
bosons can be expressed in a quite simple form. 
And, if its interpretation is based on the algebraic viewpoint, the 
mathematical structure of the orthogonal set is transparently 
understood in terms of the $su(2)$- and the $su(1,1)$-algebra. 
Further, we can construct the coherent state, which is closely connected with 
the above two algebras. Therefore, the deformation of the coherent state 
is performed under the direction of the $su(2)$- or $su(1,1)$-algebra 
in four kinds of bosons. We call the former and the latter deformation 
the $S$- and the $T$-type deformations, respectively. 
This is the outline of (I). Further, in (I), as an illustrative example, 
a concrete case of the $S$-type deformation was discussed. 
With the aid of this deformation, we can expect to describe thermal 
effects on the intrinsic structure related to 
the seniority number, the pairing rotation and the 
pairing vibration which are familiar to us in two-level shell model 
under the pairing interaction. 
In this note, two subjects on the deformation are discussed 
as the supplement of (I). The first is related 
with the $q$-deformation of the $su(2)$- and the $su(1,1)$-algebra 
in two kinds of boson operators, which is presented in Ref.\citen{KPPTY01-2}. 
For those discussions, let us imagine two many-body systems. 
Each is described in terms of the boson operators 
$({\hat a}_+, {\hat a}_+^*, {\hat b}_+, {\hat b}_+^*)$ and 
$({\hat a}_-, {\hat a}_-^*, {\hat b}_-, {\hat b}_-^*)$, respectively. 
We call the former and the latter the $(+)$- and $(-)$-systems, 
respectively. Therefore, each may be regarded as a system such as, 
for example, the Lipkin model. Of course, both systems interact mutually.

In the above model, we are concerned with the following operators 
appearing in the relations (I.3$\cdot$1a) and (I.3$\cdot$1c) : 
\begin{eqnarray}
& &{\hat S}_{\pm}^0 = {\hat S}_{\pm}^0(+) + {\hat S}_{\pm}^0(-) \ , 
\qquad
{\hat S}_0 = {\hat S}_0(+) + {\hat S}_0(-) \ , 
\label{1}\\
& &{\hat R}_{\pm}^0 = {\hat R}_{\pm}^0(a) + {\hat R}_{\pm}^0(b) \ , 
\qquad
{\hat R}_0 = {\hat R}_0(a) + {\hat R}_0(b) \ , \qquad\qquad\qquad\qquad
\label{2}
\end{eqnarray}
\vspace{-0.8cm}
\begin{subequations}\label{3}
\begin{eqnarray}
& &{\hat S}_+^0(+)={\hat a}_+^*{\hat b}_+ \ , \quad
{\hat S}_-^0(+)={\hat b}_+^*{\hat a}_+ \ , \quad
{\hat S}_0(+)=({\hat a}_+^*{\hat a}_+ - {\hat b}_+^*{\hat b}_+)/2 \ , \ \ 
\label{3a}\\
& &{\hat S}_+^0(-)={\hat a}_-^*{\hat b}_- \ , \quad
{\hat S}_-^0(-)={\hat b}_-^*{\hat a}_- \ , \quad
{\hat S}_0(-)=({\hat a}_-^*{\hat a}_- - {\hat b}_-^*{\hat b}_-)/2 \ , \ \ 
\label{3b}
\end{eqnarray}
\end{subequations}
\vspace{-0.8cm}
\begin{subequations}\label{4}
\begin{eqnarray}
& &{\hat R}_+^0(a)={\hat a}_+^*{\hat a}_- \ , \quad
{\hat R}_-^0(a)={\hat a}_-^*{\hat a}_+ \ , \quad
{\hat R}_0(a)=({\hat a}_+^*{\hat a}_+ - {\hat a}_-^*{\hat a}_-)/2 \ , \ \ 
\label{4a}\\
& &{\hat R}_+^0(b)={\hat b}_+^*{\hat b}_- \ , \quad
{\hat R}_-^0(b)={\hat b}_-^*{\hat b}_+ \ , \quad
{\hat R}_0(b)=({\hat b}_+^*{\hat b}_+ - {\hat b}_-^*{\hat b}_-)/2 \ . \ \ 
\label{4b}
\end{eqnarray}
\end{subequations}
We can see that $(\ {\hat S}_{\pm}^0(+) \ , \ {\hat S}_0(+)\ )$, 
$(\ {\hat S}_{\pm}^0(-) \ , \ {\hat S}_0(-)\ )$, 
$(\ {\hat R}_{\pm}^0(a) \ , \ {\hat R}_0(a)\ )$ and 
$(\ {\hat R}_{\pm}^0(b) \ , \ {\hat R}_0(b)\ )$ obey the $su(2)$-algebras, 
respectively. They are identical to the Schwinger boson representations. 
However, they are not mutually independent. The sets 
$({\hat S}_\pm^0 ,  {\hat S}_0)$ and $({\hat R}_\pm^0 , {\hat R}_0)$ 
obey also the $su(2)$-algebras and they are mutually independent. 
In this note, we do not contact with the $su(1,1)$-algebra. 
In associating with the above four sets, we define the following operators : 
\begin{subequations}\label{5}
\begin{eqnarray}
& &
{\hat S}(+)=({\hat a}_+^*{\hat a}_+ + {\hat b}_+^*{\hat b}_+)/2 \ ,  
\label{5a}\\
& &
{\hat S}(-)=({\hat a}_-^*{\hat a}_- + {\hat b}_-^*{\hat b}_-)/2 \ , 
\label{5b}
\end{eqnarray}
\end{subequations}
\vspace{-0.8cm}
\begin{subequations}\label{6}
\begin{eqnarray}
& &
{\hat R}(a)=({\hat a}_+^*{\hat a}_+ + {\hat a}_-^*{\hat a}_-)/2 \ ,  
\label{6a}\\
& &
{\hat R}(b)=({\hat b}_+^*{\hat b}_+ + {\hat b}_-^*{\hat b}_-)/2 \ . 
\label{6b}
\end{eqnarray}
\end{subequations}
We can regard the sets $({\hat S}_\pm^0(+) , {\hat S}_0(+))$ and 
$({\hat S}_\pm^0(-) , {\hat S}_0(-))$ as the building blocks of the 
$(+)$- and the $(-)$-systems, respectively, appearing in the Lipkin model. 
On the other hand, ${\hat R}_\pm^0(a)$ and ${\hat R}_\pm^0(b)$ 
describe the transfer of the bosons between the $(+)$- and the $(-)$-systems.

Our deformation is characterized in terms of the functions 
$(d_S({\hat I}), e_S({\hat J}), f_S({\hat S}+{\hat S}_0), 
g_S({\hat S}-{\hat S}_0))$ and 
$(d_T({\hat K}-1/2), e_T({\hat L}-1/2), f_T({\hat T}_0-{\hat T}), 
g_T({\hat T}_0+{\hat T}-2))$ appearing in the relations (I.4$\cdot$12a) 
and (I.4$\cdot$12b). In this note, we take up the case 
$f_S=g_S=f_T=g_T=1$. Then, for the $S$-type deformation, we have 
\begin{subequations}\label{7}
\begin{eqnarray}
& &{\hat S}_+^S={\hat S}_+^0 \ , \label{7a}\\
& &{\hat R}_+^S=d_S({\hat I})^{-1}d_S({\hat I}-1/2)e_S({\hat J})^{-1}
e_S({\hat J}+1/2){\hat R}_+^0 \ . 
\label{7b}
\end{eqnarray}
\end{subequations}
The $T$-type deformation gives us 
\begin{subequations}\label{8}
\begin{eqnarray}
& &{\hat S}_+^T=d_T({\hat K}-1/2)^{-1}d_T({\hat K})e_T({\hat L}-1/2)^{-1}
e_T({\hat L}){\hat S}_+^0 \ , \label{8a}\\
& &{\hat R}_+^T=d_T({\hat K}-1/2)^{-1}d_T({\hat K})e_T({\hat L}-1/2)^{-1}
e_T({\hat L}-1){\hat R}_+^0 \ . 
\label{8b}
\end{eqnarray}
\end{subequations}
The forms (\ref{7}) and (\ref{8}) are derived by the relations 
(I.5$\cdot$7), (I.5$\cdot$10), (I.5$\cdot$19) and (I.5$\cdot$22). 
The operators ${\hat I}$, ${\hat J}$, ${\hat K}$ and ${\hat L}$ 
are given in the relation (I.4$\cdot$1) : 
\begin{subequations}\label{9}
\begin{eqnarray}
& &{\hat I}=({\hat b}_+^*{\hat b}_+ + {\hat a}_+^*{\hat a}_+)/2 \ , \qquad
{\hat J}=({\hat b}_-^*{\hat b}_- + {\hat a}_-^*{\hat a}_-)/2 \ , 
\label{9a}\\
& &{\hat K}=({\hat b}_-^*{\hat b}_- - {\hat a}_+^*{\hat a}_+ +1)/2 \ , \qquad
{\hat L}=({\hat b}_+^*{\hat b}_+ - {\hat a}_-^*{\hat a}_- +1)/2 \ .\quad 
\label{9b}
\end{eqnarray}
\end{subequations}
As was shown in the relation (I.4$\cdot$13), $d_S$, $e_S$, $d_T$ and $e_T$ 
should obey 
\begin{subequations}\label{10}
\begin{eqnarray}
& &d_S(1/2)=d_S(0) \ , \qquad e_S(1/2)=e_S(0) \ , 
\label{10a}\\
& &d_T(1/2)=d_T(0) \ , \qquad e_T(1/2)=e_T(0) \ . 
\label{10b}
\end{eqnarray}
\end{subequations}
The above is the basic framework of the present note.

First, we will discuss the first subject which was not mentioned in (I). 
In Ref.\citen{KPPTY01-2}, we treated the $su(2)_q$- and 
the $su(1,1)_q$-algebra in two kinds of boson operators. 
Especially, three forms based on the use of the parameter $q$ 
were shown including the most popular form. 
Therefore, it may be interesting to investigate which forms 
appear in the case of four kinds of boson operators. 
For the $S$-type deformation, we set up the relation 
\begin{equation}\label{15}
d_S({\hat I}+1/2)^{-1}d_S({\hat I})=(q_S)^{{\hat I}} \ , \qquad
e_S({\hat J})^{-1}e_S({\hat J}+1/2)=(q_S)^{{\hat J}} \ .
\end{equation}
Here, $q_S$ denotes a positive parameter. Substituting the form (\ref{15}) 
into the relation (\ref{7b}), we have 
\begin{eqnarray}
& &{\hat R}_+^S={\hat R}_+^S(a) \cdot (q_S)^{{\hat R}(b)}
+{\hat R}_+^S(b) \cdot (q_S)^{{\hat R}(a)} \ , 
\label{16}\\
& &{\hat R}_+^S(a)=(q_S)^{{\hat R}(a)-{1}/{2}}
{\hat R}_+^0(a) \ , \qquad 
{\hat R}_+^S(b)=(q_S)^{{\hat R}(b)-{1}/{2}}{\hat R}_+^0(b) \ . 
\label{17}
\end{eqnarray}
The form (\ref{17}) is the same as that\cite{KPPTY01-2} 
obtained in terms of the $q$-deformation by Penson and 
Solomon.\cite{PS99} 
As is clear from the form (\ref{2}), ${\hat R}_+^0$ is a superposition of 
${\hat R}_+^0(a)$ and ${\hat R}_+^0(b)$ with the equal weights. 
In the case of ${\hat R}_+^S$, the weights are generally different 
from each other. 
Further, we can show that $[2{\hat R}_0]_S$ is also expressed in the form 
\begin{eqnarray}
& &[2{\hat R}_0]_S=[2{\hat R}_0(a)]_S\cdot (q_S)^{2{\hat R}(b)}
+[2{\hat R}_0(b)]_S\cdot (q_S)^{2{\hat R}(a)} \ , 
\label{add-5-1}
\end{eqnarray}
\vspace{-0.85cm}
\begin{subequations}\label{add-5-2}
\begin{eqnarray}
& &[2{\hat R}_0(a)]_S=(q_S)^{2({\hat R}(a)-1/2)} (2{\hat R}_0(a)) \ , 
\qquad\qquad\qquad\ 
\label{add-5-2a}\\
& &[2{\hat R}_0(b)]_S=(q_S)^{2({\hat R}(b)-1/2)} (2{\hat R}_0(b)) \ . 
\label{add-5-2b}
\end{eqnarray}
\end{subequations}
The proof is omitted. The form of the superposition is similar to 
that of the relation (\ref{16}). 
We are able to obtain the results for the $T$-type 
deformation similar to the form (\ref{17}). Let us start from the 
following relation : 
\begin{equation}\label{18-0}
d_T({\hat K}-1/2)^{-1}d_T({\hat K})=(q_T)^{-({\hat K}-1/2)} \ , \qquad
e_T({\hat L})^{-1}e_T({\hat L}-1/2)=(q_T)^{-({\hat L}-1/2)} \ .
\end{equation}
Here, of course, $q_T$ denotes a positive parameter. 
Then, we have 
\begin{subequations}\label{18}
\begin{eqnarray}
& &{\hat S}_+^T={\hat S}_+^T(+)\cdot (q_T)^{-({\hat S}(-)-1/2)}
+{\hat S}_+^T(-) \cdot (q_T)^{{\hat S}(+)-1/2} \ , 
\label{18a}\\
& &{\hat R}_+^T={\hat R}_+^T(a)\cdot (q_T)^{-({\hat R}(b)-1/2)}
+{\hat R}_+^T(b) \cdot (q_T)^{{\hat R}(a)-1/2} \ , 
\qquad\qquad\qquad\qquad\qquad\qquad
\label{18b}
\end{eqnarray}
\end{subequations}
\vspace{-0.8cm}
\begin{subequations}\label{19}
\begin{eqnarray}
& &{\hat S}_+^T(+)=(q_T)^{{\hat S}(+)-1/2}
{\hat S}_+^0(+) \ , \qquad 
{\hat S}_+^T(-)=(q_T)^{-({\hat S}(-)-{1}/{2})}{\hat S}_+^0(-) \ ,  \quad
\label{19a}\\
& &{\hat R}_+^T(a)=q_T^{1/2}\cdot(q_T)^{{\hat R}(a)-1/2}
{\hat R}_+^0(a) \ , \quad 
{\hat R}_+^T(b)=q_T^{1/2}\cdot
(q_T)^{-({\hat R}(b)-{1}/{2})}{\hat R}_+^0(b) \ .  \qquad\ \ 
\label{19b}
\end{eqnarray}
\end{subequations}
In a form similar to the case of the $S$-type deformation, 
${\hat S}_+^T$ and ${\hat R}_+^T$ can be expressed in terms of linear 
combinations for $({\hat S}_+^T(+) , {\hat S}_+^T(-))$ and 
$({\hat R}_+^T(a) , {\hat R}_+^T(b))$, respectively. 
However, the $q_T$-dependence of 
${\hat S}_+^T(-)$ and ${\hat R}_+^T(b)$ are different from that of 
${\hat S}_+^T(+)$ and ${\hat R}_+^T(a)$, respectively. 
The weights of the linear combinations are also different from 
those for the $S$-type deformation. 
In this case, also we can show that $[2{\hat S}_0]_T$ and $[2{\hat R}_0]_T$ 
are of the following forms : 
\begin{subequations}\label{add-6-1}
\begin{eqnarray}
& &[2{\hat S}_0]_T=[2{\hat S}_0(+)]_T\cdot (q_T^{-1})^{2({\hat S}(-)-1/2)}
+[2{\hat S}_0(-)]_T\cdot (q_T)^{2({\hat S}(+)-1/2)} \ , 
\label{add-6-1a}\\
& &[2{\hat R}_0]_T=[2{\hat R}_0(a)]_T\cdot (q_T^{-1})^{2({\hat R}(b)-1/2)}
+[2{\hat R}_0(b)]_T\cdot (q_T)^{2({\hat R}(a)-1/2)} \ , 
\label{add-6-1b}
\end{eqnarray}
\end{subequations}
\vspace{-0.8cm}
\begin{subequations}\label{add-6-2}
\begin{eqnarray}
& &[2{\hat S}_0(+)]_T=(q_T)^{2({\hat S}(+)-1/2)}(2{\hat S}_0(+)) \ , 
\nonumber\\
& &
[2{\hat S}_0(-)]_T=(q_T^{-1})^{2({\hat S}(-)-1/2)}(2{\hat S}_0(-)) \ , 
\qquad\qquad\qquad\qquad\qquad\qquad
\label{add-6-2a}\\
& &[2{\hat R}_0(a)]_T=q_T\cdot
(q_T)^{2({\hat R}(a)-1/2)}(2{\hat R}_0(a)) \ , 
\nonumber\\
& &
[2{\hat R}_0(b)]_T=q_T\cdot
(q_T^{-1})^{2({\hat R}(b)-1/2)}(2{\hat R}_0(b)) \ . 
\label{add-6-2b}
\end{eqnarray}
\end{subequations}
In the case of the most popular 
deformation $(q^n-q^{-n})/(q-q^{-1})$, it is impossible to 
derive simple forms such as the above.

Next, let us discuss the second subject, which may be more 
interesting than the first. 
In (I), we investigated the following case for the relation (\ref{7b}) : 
\begin{equation}\label{11}
d_S({\hat I}+1/2)^{-1}d_S({\hat I})
=\sqrt{1-2{\hat I}/Z_S} \ , \quad
e_S({\hat J})^{-1}e_S({\hat J}+1/2)=\sqrt{1-2{\hat J}/Z_S} \ . 
\end{equation}
Then, ${\hat R}_+^S$ can be expressed in the form 
\begin{eqnarray}\label{12}
{\hat R}_+^S&=&
Z_S^{-1}\Bigl[
{\hat a}_+^*\sqrt{(Z_S-{\hat b}_+^*{\hat b}_+)-{\hat a}_+^*{\hat a}_+}
\cdot\sqrt{(Z_S-{\hat b}_-^*{\hat b}_-)-{\hat a}_-^*{\hat a}_-}\ {\hat a}_- 
\nonumber\\
& &\qquad
+{\hat b}_+^*\sqrt{(Z_S-{\hat a}_+^*{\hat a}_+)-{\hat b}_+^*{\hat b}_+}
\cdot\sqrt{(Z_S-{\hat a}_-^*{\hat a}_-)-{\hat b}_-^*{\hat b}_-}\ {\hat b}_- 
\Bigl] \ . 
\end{eqnarray}
Of course, $Z_S$ denotes a positive parameter characterizing the present 
model, and in (I) it was assumed to be large. 
As a possible interpretation for the above expression, 
we mentioned in (I) that, for example, 
${\hat a}_+^*\sqrt{(Z_S-{\hat b}_+^*{\hat b}_+)-{\hat a}_+^*{\hat a}_+}$ 
is nothing but the Holstein-Primakoff representation of the 
$su(2)$-spin with the magnitude $(Z_S-{\hat b}_+^*{\hat b}_+)/2$. 
On the basis of the above fact, we showed an illustrative example in (I) 
for the pairing correlation in two-level shell model. 
Corresponding to the case described in (\ref{11}), we adopt the 
following form for the $T$-type deformation : 
\begin{equation}\label{13}
d_T({\hat K}-1/2)^{-1}d_T({\hat K})
=\sqrt{1+\frac{2{\hat K}-1}{Z_T+1}} \ , \quad
e_T({\hat L})^{-1}e_T({\hat L}-1/2)=\sqrt{1+\frac{2{\hat L}-1}{Z_T+1}} \ . 
\end{equation}
Then, ${\hat S}_+^T$ and ${\hat R}_+^T$ can be expressed as 
\begin{subequations}\label{14}
\begin{eqnarray}
& &{\hat S}_+^T=
{\hat S}_+^0\sqrt{\frac{Z_T-{\hat a}_+^*{\hat a}_+ + {\hat b}_-^*{\hat b}_-}
{Z_T-{\hat a}_-^*{\hat a}_- + {\hat b}_+^*{\hat b}_+}} \ , 
\label{14a}\\
& &{\hat R}_+^T
=(Z_T+1)^{-1}
\Bigl[
{\hat a}_+^*\sqrt{(Z_T+{\hat b}_-^*{\hat b}_-)-{\hat a}_+^*{\hat a}_+}
\cdot
\sqrt{(Z_T+{\hat b}_+^*{\hat b}_+)-{\hat a}_-^*{\hat a}_-}\ {\hat a}_- 
\nonumber\\
& &\qquad\qquad\qquad\qquad
+{\hat b}_+^*\sqrt{(Z_T+1-{\hat a}_-^*{\hat a}_-)+{\hat b}_+^*{\hat b}_+}
\cdot
\sqrt{(Z_T+1-{\hat a}_+^*{\hat a}_+)+{\hat b}_-^*{\hat b}_-}\ {\hat b}_-
\Bigl] \ . \ \ \nonumber\\
& &
\label{14b}
\end{eqnarray}
\end{subequations}
Here, $Z_T$ denotes a positive parameter characterizing the present model.

Contrasting with the form (\ref{12}), let us give a possible 
interpretation of the relation (\ref{14}). As is clear from the form 
(\ref{14a}), ${\hat S}_+^T$ is different from ${\hat S}_+^0$ in 
contrast to the case of ${\hat S}_+^S$. 
However, in the region where the magnitude of the $su(2)$-spin for 
the $(+)$-system $({\hat S}(+)=({\hat a}_+^*{\hat a}_+ + {\hat b}_+^*
{\hat b}_+)/2)$) is nearly equal to that for the $(-)$-system 
$({\hat S}(-)=({\hat a}_-^*{\hat a}_- + {\hat b}_-^*
{\hat b}_-)/2)$), the behavior of ${\hat S}_+^T$ is almost the 
same as that of ${\hat S}_+^0$. 
In the case of ${\hat R}_+^T$, new aspect appears. For example, we see 
that ${\hat a}_+^*\sqrt{(Z_T+{\hat b}_-^*{\hat b}_-)-{\hat a}_+^*{\hat a}_+}$ 
is nothing but the Holstein-Primakoff representation of the $su(2)$-spin, 
which is in the same situation as that of ${\hat R}_+^S$. 
Then, qualitatively, the behavior of the first term of ${\hat R}_+^T$ 
may be expected to be similar to that of ${\hat R}_+^S$. 
However, the behavior of the second term is different from that of 
${\hat R}_+^S$. For example, 
${\hat b}_+^*\sqrt{(Z_T+1-{\hat a}_-^*{\hat a}_-)+{\hat b}_+^*{\hat b}_+}$ 
is nothing but the Holstein-Primakoff representation of 
the $su(1,1)$-algebra with the magnitude of 
$(Z_T+1-{\hat a}_-^*{\hat a}_-)/2$. 
However, if the magnitude of ${\hat R}(b)\ (=({\hat b}_+^*{\hat b}_+
+{\hat b}_-^*{\hat b}_-)/2)$ is nearly equal to $(Z_S-Z_T)/2$, 
the behaviors of the first and the second term are almost the same as 
those of ${\hat R}_+^S$. Therefore, except for the same special cases, 
the behavior of the transfer between the $(+)$- and $(-)$-systems 
in the $T$-type deformation is different from that in the $S$-type 
deformation. 
Of course, the case of ${\hat S}_+^T$ is also in the same situation as 
the case of ${\hat S}_+^S$.

In the above, we discussed two subjects which were not mentioned in (I). 
The first is rather formal and it may be helpful for understanding 
of theoretical structure of the deformed boson scheme proposed by the 
present authors. But, the second may be expected to be a possible 
approach to describe various aspects of many-body systems related to 
the $su(2)$- and the $su(1,1)$-algebra. These investigations are our 
further problems.



\vspace{-0.2cm}










\begin{thebibliography}{99}
\bibitem{KPPTYsub}
A. Kuriyama, C. Provid\^encia, J. da Provid\^encia, Y. Tsue and M.~Yamamura, 
        to appear in Prog.~Theor.~Phys. {\bf 108} (2002), No.2.
\bibitem{KPPTY01-1}
A. Kuriyama, C. Provid\^encia, J. da Provid\^encia, Y. Tsue and M.~Yamamura, 
        Prog. Theor. Phys. {\bf 106} (2001), 751.
\bibitem{KPPTY01-2}
A. Kuriyama, C. Provid\^encia, J. da Provid\^encia, Y. Tsue and M.~Yamamura, 
        Prog. Theor. Phys. {\bf 106} (2001), 765. 
\bibitem{KPPTY02}
A. Kuriyama, C. Provid\^encia, J. da Provid\^encia, Y. Tsue and M.~Yamamura, 
        Prog. Theor. Phys. {\bf 107} (2002), 65.
\bibitem{KPY00}
A. Kuriyama, J. da Provid\^encia and M.~Yamamura, 
        Prog. Theor. Phys. {\bf 103} (2000), 305.
\bibitem{PS99}
K. A. Penson and A. I. Solomon, J. Math. Phys. {\bf 40} (1999), 
2354.
\end{thebibliography}
\end{document}